\documentclass[twocolumn,letterpaper]{revtex4-1}
\usepackage{graphics}
\usepackage{graphicx}
\usepackage{epsfig}
\usepackage{units}
\begin{document}
\title{Excitations of strange bottom baryons}
\author{R. M. Woloshyn}                     
\affiliation{TRIUMF, 4004 Wesbrook Mall, Vancouver, British Columbia V6T 2A3, Canada }
\begin{abstract}
The ground state and first excited state masses of 
$\Omega_{b}$  and $\Omega_{bb}$ baryons are calculated in lattice QCD using 
dynamical $2+1$ flavour gauge fields. A set of baryon operators employing
different combinations of smeared quark fields was used in the framework
of the variational method. Results for radial excitation energies were
confirmed by carrying out a supplementary multiexponential fitting analysis.
Comparison is made with quark model calculations.
\end{abstract}
\maketitle
\section{Introduction}
\label{intro}
The calculation of hadronic excitation energies is an important 
area of activity for lattice QCD. Hadrons containing both heavy and
light quarks present
an interesting challenge since one has to deal with quarks at
many different mass scales. Recently results for radial excitations
of heavy-light hadrons were reviewed \cite{Woloshyn:2016cgd} 
and it was found that in the 
meson sector there was reasonably good agreement between lattice QCD 
simulations, quark models and available (but still incomplete) experimental 
data. However, for baryons containing a single charm or bottom quark there
seems to be a puzzle. The few results obtained for radial excitations 
of S-wave singly heavy baryons are consistently larger than quark model
values. Whereas the quark model suggests that excitation energies of 
heavy-light baryons should be smaller than that of heavy-light mesons the 
lattice results reviewed in \cite{Woloshyn:2016cgd} do not show this 
qualitative feature.  Unfortunately, experiment is unlikely to settle 
this issue any time soon.

In addition to predicting that baryonic excitations are smaller than 
mesonic ones, quark models also predict that the lowest lying radial
excitation of a doubly heavy baryon should be significantly smaller 
than of a singly heavy baryon (see, for example, fig. 3 in \cite{Roberts:2007ni}). 
This is due
to the different excitation modes at play \cite{Roberts:2007ni,Ebert:2002ig,Yoshida:2015tia}. 
In a singly heavy baryon the
lowest lying excitation is due to the motion of the heavy quark relative to
the diquark system formed by the light quarks. In a doubly heavy baryon 
it is the excitation of the heavy quark pair that is important. 
This effect is particularly pronounced for bottom baryons. It is
natural to ask if the features of the spectrum which these quark model
mechanisms predict are reproduced in a lattice QCD simulation. This is
the theme of the present paper.

In this work we focus on the lattice simulation of spin $\nicefrac{1}{2}$ 
baryons $\Omega_{b}$  and $\Omega_{bb}.$ Since the valence quark content
of these baryons consists only of strange and bottom the values of the
$u$ and $d$ sea quark masses are of secondary importance. For our simulation 
the $u$ and $d$ quarks are near physical corresponding to a pion mass 
of $156(7) MeV$ so $u,d$ mass extrapolation is not carried out. Secondly,
with strange quark valence content statistical fluctuations are considerably
smaller than what would have been possible if $u,d$ valence quarks would
have been used. 

The lattice setup for the present work is described in sect. \ref{sec:1}. 
In sect. \ref{sec:2} the analysis method and results are presented. The
results are compared to a number of recent quark model calculations. 
A summary is given in sect. \ref{sec:3}.

\section{Lattice setup}
\label{sec:1}
The $2+1$ dynamical flavour gauge field configurations used in this 
work were generated by the PACS-CS Collaboration \cite{pacscs09}
on a $32^4 \times 64$ lattice using the Iwasaki action
($\beta = 1.90$) for the gauge field and the clover-Wilson action for the
fermions. The strange quark hopping parameter used here was $0.13666$ which is 
slightly different than that used by the PACS-CS Collaboration and is in line
with the value determined from earlier work on the $D_{s}$ spectrum 
\cite{Lang:2014yfa}. The strange
quark clover coefficient was $1.715$. The gauge field ensemble had 198 
configurations. For this ensemble PACS-CS determined a lattice spacing 
of $a = 0.0907(13) fm$. 

The bottom quark was described using tadpole-improved lattice NRQCD \cite{Lepage:1992}. The
Hamiltonian is the same as used in previous calculation of bottom baryon
masses and may be found in the Appendix of ref. \cite{Lewis:2009}
Terms up to order $v^4$ were
retained in the nonrelativistic expansion. The $b$-quark bare mass was 
$1.945$ as determined in ref. \cite{Lewis:2011} from tuning to S-wave 
bottomonium. The average
link in Landau gauge, estimated to be $0.8463$, was used for tadpole improvement
and the stability parameter $n$ appearing in the NRQCD Hamiltionian was taken
to be 4. As a check of the lattice NRQCD setup the $\Upsilon$ - $\eta_b$ mass
difference was calculated with the result $58.1(1.5) MeV$ in excellent
agreement with the experimental value $57.9(2.4) MeV$ obtained using the 
PDG \cite{pdg} value for the $\Upsilon$ mass and the Belle result \cite{Belle:2012}
for $\eta_b$. 

The baryon operators used in our calculation take the form 
\begin{equation}
\epsilon^{abc}[q_{a}^{T}C\gamma_{5}q'_{b}]q_{c}
\label{eq:bary}
\end{equation}
where for the spin $\nicefrac{1}{2}$ $\Omega_{b}$ baryon $q$ is a 
strange quark field and $q'$ is bottom.
For $\Omega_{bb}$, $q$ is a bottom quark field and $q'$ is strange.

The simplest operator that one can construct of the form (\ref{eq:bary})
has all quark fields at the same space-time point. However to disentangle
ground and excited state contributions to correlation functions it is
advantageous to use a variety of spatially smeared operators which will
lead to different admixtures of states in the correlators. Successful 
phenomenological descriptions of baryons are often made using a quark-diquark
model which implies strong spatial correlations among the constituents.
Ideally, we would like to construct lattice baryon operators incorporating 
quark-quark correlations. However a usable scheme for doing this is not
available so here independent spatial smearing of the quark fields is
carried out. 

Quark fields are smeared according to 
\begin{equation}
\tilde{\psi}({\bf x})=\sum_{\bf y} f({\bf x}-{\bf y})\,\psi({\bf y}) \label{gaugefixsmear}
\end{equation}
where f is a gauge field independent profile function. Since this is not gauge 
covariant a gauge fixing to Coulomb gauge is carried out on the gauge field
links prior to use. A choice for the profile $f$ which has been used
successfully in NRQCD applications is motivated by the shape of wavefunctions
for the Coulomb potential \cite{Gray:2005,Davies:2010}. 
In this work we use mostly a smearing function of 
the form  $e^{-\frac{r}{a_{0}}}$ where $r$ is the shortest distance between 
${\bf y}$  and ${\bf x}$ in a periodic box. This will be referred to as ground 
state smearing. An excited state smearing function  
$(a_n - r)e^{-\frac{r}{a_{1}}})$ was also considered. Since excited state
smearing tends to lead to noisier correlators this smearing was used only 
in a supplementary calculation.

The strange quark field was given a more spatially extended profile than the 
bottom quark. After some trials the smearing parameters (in lattice units) chosen for this 
work were $a_0 = 1.6, a_1 = 3.2, a_n = 3.17$ for the bottom quark and
$a_0 = 3.0, a_1 = 5.5, a_n = 7.0$ for strange.
 
\section{Analysis and results}
\label{sec:2}

\begin{figure}
\centerline{
\includegraphics*[width=90mm]{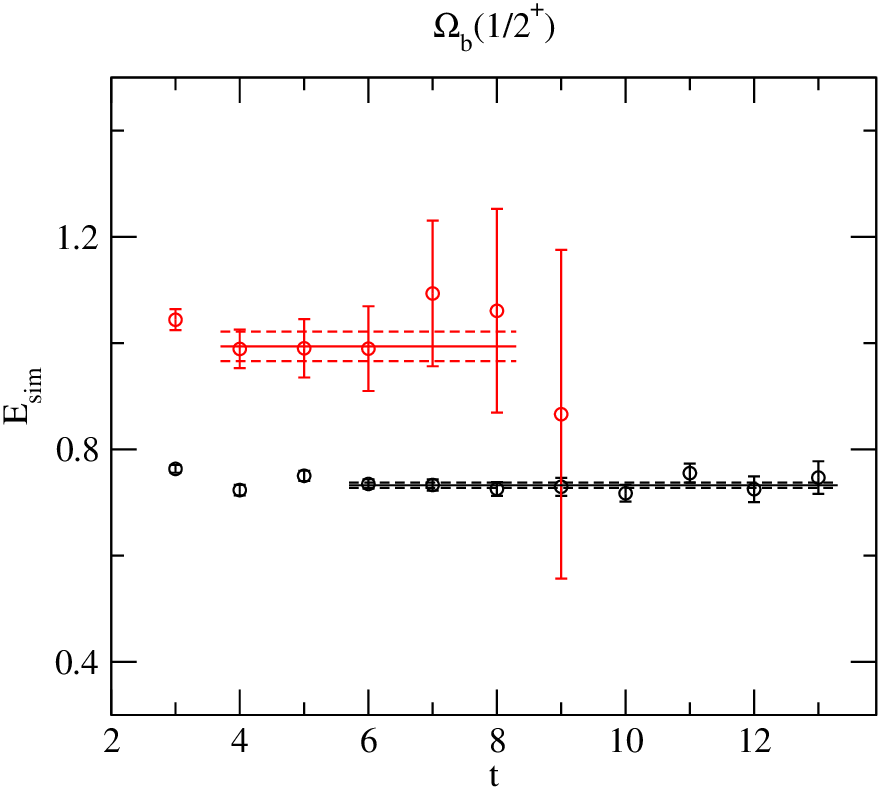}}
\caption{Effective simulation energies in lattice units for the ground and 
first excited state of $\Omega_{b}$. The horizontal lines indicate the fit
value and time range used.}
\label{fig:1}
\end{figure}

Lattice Euclidean-time correlation functions are typically dominated by
the ground state after a fairly small number of time steps. To extract 
excited state information requires some work. The approach used here is
the so-called variational method \cite{Luscher:1990,Michael:1985,Blossier:2009}. 
A set of basis operators $\{O\}$ is chosen 
and one constructs the correlator matrix 
\begin{equation}
C_{ij}(t=t_f-t_i)=\langle 0|O_i(t_f)O_j^\dagger(t_i)|0\rangle
\end{equation}
where $t_i$ and $t_f$ denote the source and sink times. The generalized
eigenvalue problem \cite{Blossier:2009} is solved for each time step larger than some reference
time $t_0$
\begin{equation}
C(t)\vec{w}^{(k)}=\lambda^{(k)}(t)C(t_0)\vec{w}^{(k)}.
\label{eq:gevp}
\end{equation}
Energies are extracted from the time dependence of the eigenvalues $\lambda$
and for a well chosen basis and $t_0$ the eigenvalues of the lowest lying 
states are dominated by a single exponential function.

For the main calculation a set of six operators with different combinations
of unsmeared ($p$) and ground state smeared ($g$) fields is used. The operators 
are denoted as 
\begin{equation}
ppp, ppg, gpg, pgp, pgg, ggg 
\label{eq:ops1}
\end{equation}
where the first and third
letters indicate the smearing level of $q$ fields in eq. (1) (strange for
$\Omega_b$, bottom for  $\Omega_{bb}$) and the second letter indicates the 
smearing level of $q'$ field (bottom for $\Omega_b$, strange for $\Omega_{bb}$).
Note that for $\Omega_b$ energies of both positive and negative parity states 
can be extracted by choosing different components of the relativistic operator.

\begin{figure}
\centerline{
\includegraphics*[width=90mm]{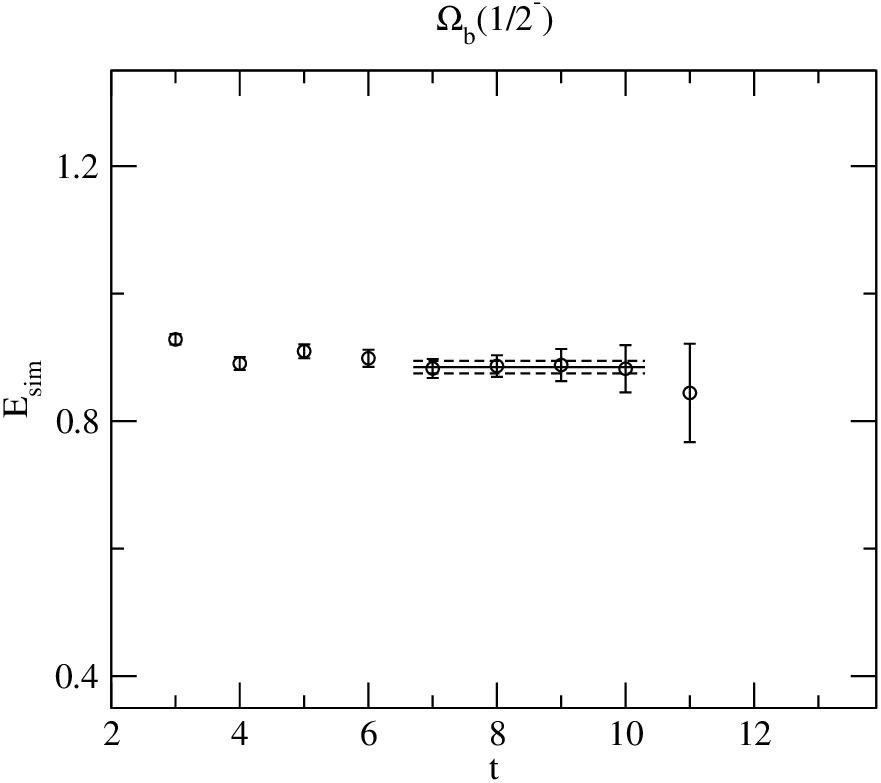}}
\caption{Effective simulation energies in lattice units for the ground state of 
the negative parity $\Omega_{b}$. The horizontal lines indicate the fit
value and time range used.}
\label{figs:2}
\end{figure}

\begin{figure}
\centerline{
\includegraphics*[width=90mm]{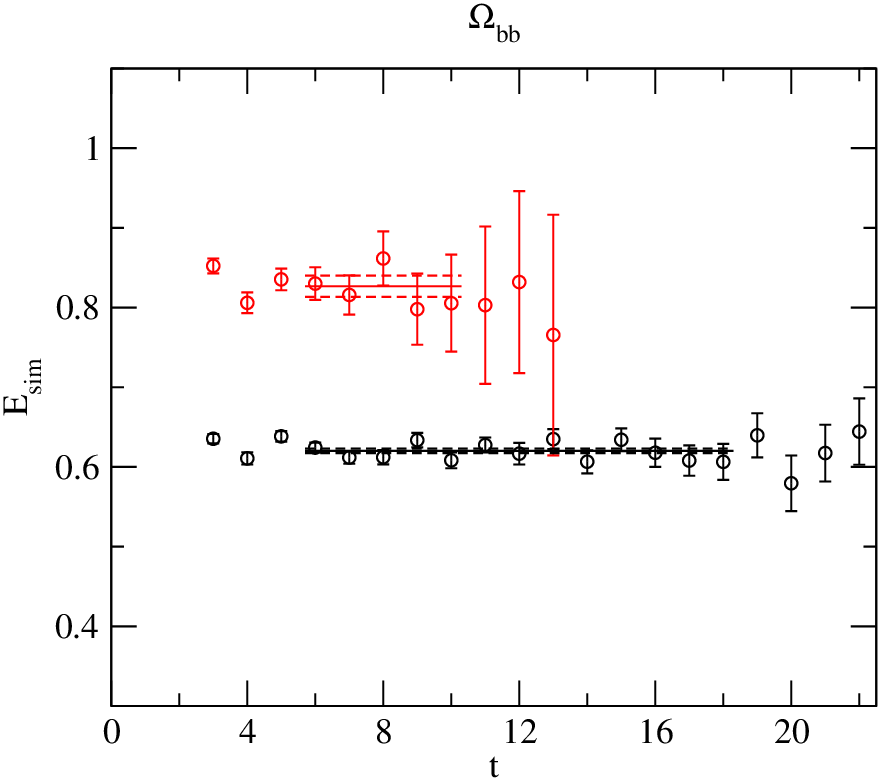}}
\caption{Effective simulation energies in lattice unitsfor the ground and 
first excited state of $\Omega_{b}$. The horizontal lines indicate the fit
value and time range used.}
\label{figs:3}
\end{figure}

For each configuration correlation functions were averaged over 16
different source time positions. The generalized eigenvalue problem
(\ref{eq:gevp}) was solved using $t_0 = 2$. The effective simulation
energies (lattice units) for the positive parity $\Omega_b$ extracted 
from the eigenvalues 
of the two lowest states are shown in fig. \ref{fig:1}. The horizontal 
lines show the fitted simulation energy values along with time range used to
determine them.
The ground state is very well determined. The first excited state is well
separated from the ground state and the energy can be extracted with a
reasonably small statistical uncertainty. Figure 2 shows the ground state
simulation energy for $\Omega_{b}({\nicefrac{1}{2}}^-)$. In this case only 
the ground state is reasonably well determined.

The simulation energy results for $\Omega_{bb}$ are plotted in fig. 3. For
$\Omega_{bb}$ only the positive parity state is available due to the
nonrelativistic treatment of the bottom quark. 

\begin{table}
\caption{Extracted simulation energies in lattice units for the variational 
and multiexponential fitting methods.}
\label{tab:1}       
\centering{}\begin{tabular}{lll}
\hline\noalign{\smallskip}
State & Variational  & Multiexponential\\
\noalign{\smallskip}\hline\noalign{\smallskip}
$\Omega_{b}({\nicefrac{1}{2}}^+)$ g.s. & 0.7324(50)(24) & 0.7262(43) \\
$\Omega_{b}({\nicefrac{1}{2}}^+)$ excited & 0.9938(278)(10) & 0.9904(329) \\
$\Omega_{b}({\nicefrac{1}{2}}^-)$ g.s. &  0.8847(99)(90) & 0.8801(44) \\
$\Omega_{bb}({\nicefrac{1}{2}}^+)$ g.s. & 0.6201(29)(20) & 0.6137(22) \\
$\Omega_{bb}({\nicefrac{1}{2}}^+)$ excited & 0.8267(133)(24) & 0.8015(139) \\
\noalign{\smallskip}\hline
\end{tabular}
\end{table}

The values for the fitted simulation energies are given in table 1. The 
first error is statistical, the second is an estimate of the uncertainty
due to different choices of time range to include in the fit.

\begin{figure}
\centerline{
\includegraphics*[width=85mm]{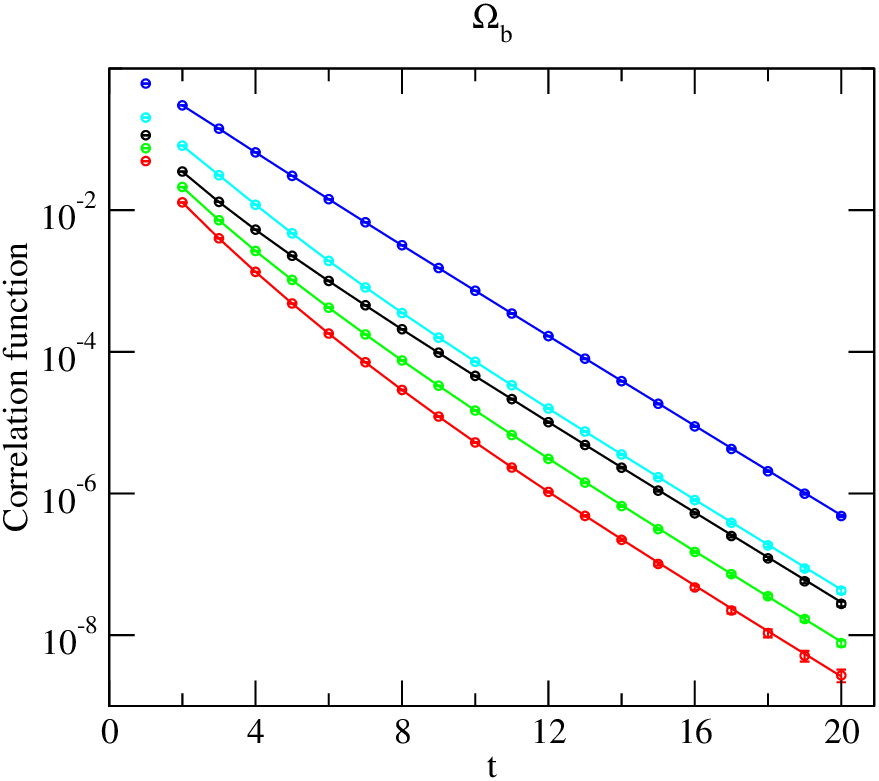}}
\caption{Correlation functions for the positive parity $\Omega_b$ along with the
result of a simultaneous four-term multiexponential fit.}
\label{figs:4}
\end{figure}

To provide additional confirmation that our extraction of excited state 
energies is reliable a second analysis was done using a different method
and a different set of correlation functions. Since smearing at the sink usually 
leads to noisier correlators than smearing at the source a set of correlation
function was constructed with source operators using combinations of fields
\begin{equation}
gpg, gpe, epe, ggg, ege  
\label{eq:ops2}
\end{equation}
where $e$ denotes smearing with an excited state profile
and a local operator $ppp$ at the sink. A simultaneous constrained 
multiexponential fit \cite{Lepage:2001ym} to the five correlation functions 
was done using four terms. As in the 
variational analysis correlation functions were averaged over a set of sixteen
different time sources for each gauge configuration. The correlation functions
and fits are shown in fig. \ref{figs:4} to fig. \ref{figs:6}. The time range 
used for fitting was 2 to 18
for $\Omega_b$ and 3 to 20 for $\Omega_{bb}$. The source corresponds to $t=1$.
The simulation energies for the ground and first excited states are given in
table \ref{tab:1}. They are compatible within statistical errors with the 
results obtained using the correlator matrix variational method.

\begin{figure}
\centerline{
\includegraphics*[width=85mm]{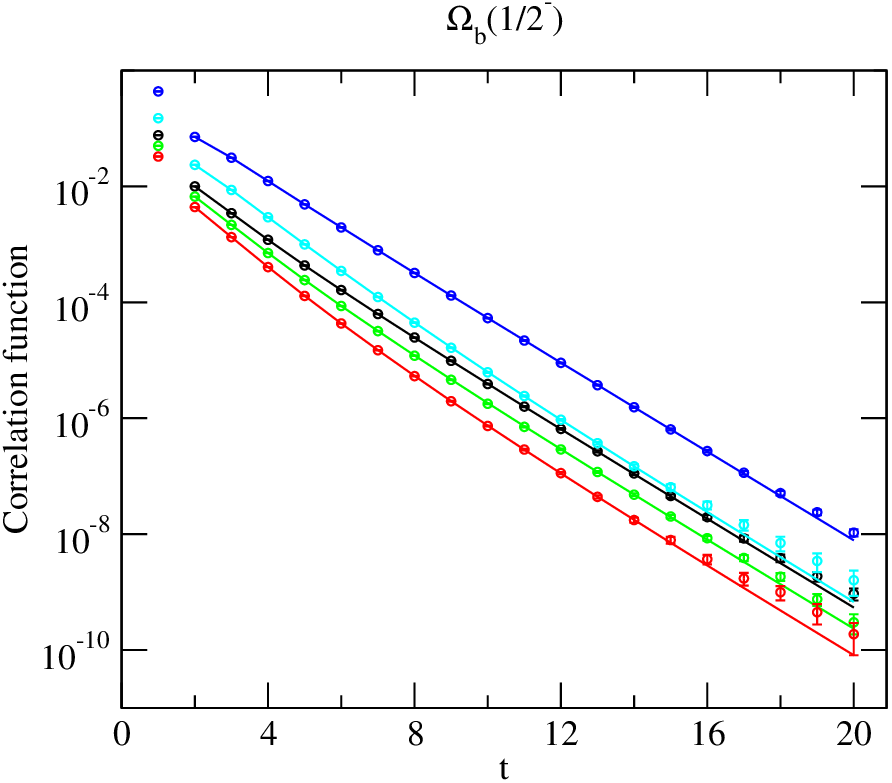}}
\caption{Correlation functions for the negative parity $\Omega_b$ along with the
result of a simultaneous four-term multiexponential fit.}
\label{figs:5}
\end{figure}

\begin{figure}
\centerline{
\includegraphics*[width=85mm]{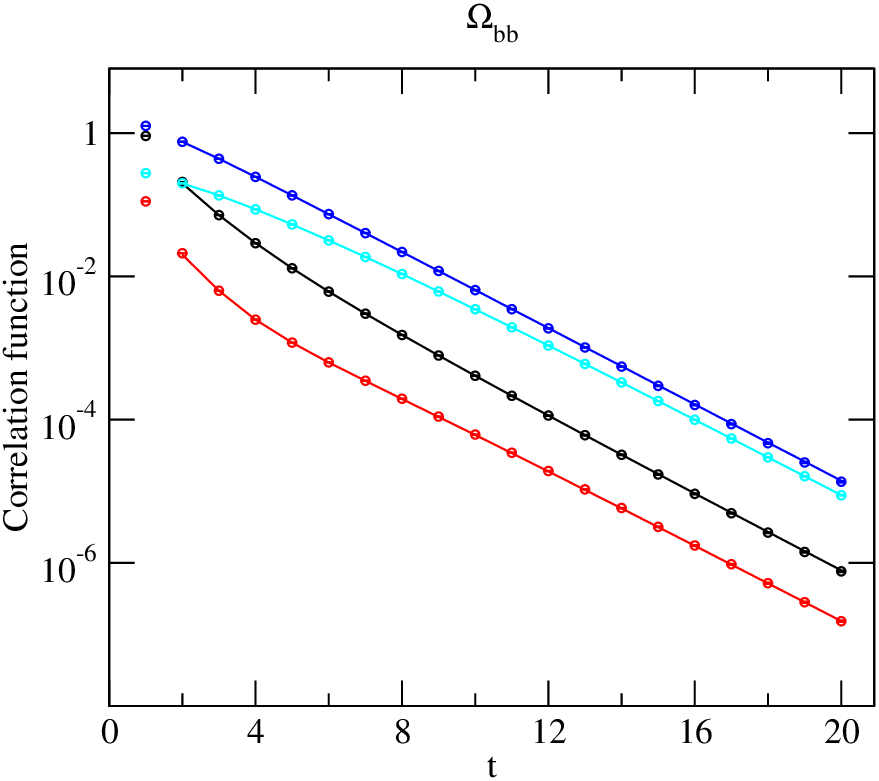}}
\caption{Correlation functions for the $\Omega_{bb}$ along with the
result of a simultaneous four-term multiexponential fit. Due to the node
in the excited state profile file function the correlator with source
smearing $gpe$ has a negative piece and is omitted from the plot.}
\label{figs:6}
\end{figure}

Since quark mass has been removed from the lattice NRQCD Hamiltonian 
the simulation energies extracted from the correlation do not give the 
hadron mass directly. However, differences between simulation energies
in lattice units are related to mass differences by the inverse of the 
lattice spacing. Our calculated mass differences ${\Delta}M$ between the 
first radial excitation and the ground state for $\Omega_b$ and for
$\Omega_{bb}$ are given in table \ref{tab:2}. The three errors shown for this 
work are the statistical error and uncertainties due to fitting time range 
and the lattice spacing determination.
Table  \ref{tab:2} also shows the calculated mass difference between the 
positive and negative parity ground states of $\Omega_b$. For comparison we
note that an earlier lattice study \cite{Woloshyn:2016cgd}
gave mass differences for $\Lambda_b$
and $\Sigma_b$ of $344(105) MeV$ and $252(60) MeV$ respectively. The PGD
value for the $\Lambda_b({\nicefrac{1}{2}}^-) - \Lambda_b({\nicefrac{1}{2}}^+)$ 
mass difference is $293(1) MeV$.

\begin{table*}
\caption{Mass differences and masses in MeV from this work and some
recent quark model calculations. The errors associated to this work are due
to statistics, fitting and scale setting}
\label{tab:2}       
\centering{}\begin{tabular}{llllll}
\hline\noalign{\smallskip}
& This work  & ref. \cite{Roberts:2007ni} & ref. \cite{Ebert:2002ig,Ebert:2011kk}& ref. \cite{Yoshida:2015tia} & ref. \cite{Garcilazo:2007eh}\\
\noalign{\smallskip}\hline\noalign{\smallskip}
${\Delta}M(\Omega_b)$ & $569(61)(6)(8)$ & $391$ & $386$ & $441$ & $330$\\
${\Delta}M(\Omega_{bb})$ & $450(29)(6)(6)$ & $239$ & $251$ & $260$ & \\
$\Omega_b({\nicefrac{1}{2}}^-) - \Omega_b({\nicefrac{1}{2}}^+)$ & $331(24)(19)(5)$ & $220$ & $275$ & $257$ & $241$ \\
$M_{\Omega_b}$ & $6038(11)(5)(18)$ & $6081$ & $6064$ & $6076$ & $6037$\\
$M_{\Omega_{bb}}$ & $10238(6)(4)(11)$ & $10454$ & $10359$ & $10447$ &  \\
\noalign{\smallskip}\hline
\end{tabular}
\end{table*}

One way to get the baryon mass would be to calculate the kinetic energy
as a function of momentum and determine a kinetic mass. However, this would
be very time consuming and likely would have large statistical errors. Instead 
we use the fact the bottom quark bare mass was tuned by fitting the mass of 
bottomonium. The simulation energy of $\Upsilon$ can related to the mass
by 
\begin{equation}
M_{\Upsilon}=E_{sim}^{\Upsilon}+{\frac{1}{2}}(ZM_0-E_{shift}).
\label{eq:mups}
\end{equation}
In (\ref{eq:mups}) $M_0$ is the bottom bare mass, $Z$ is the mass
renormalization factor and $E_{shift}$ is an additive mass shift that
appears in lattice NRQCD \cite{Lepage:1992}.  
These quantities are independent of the hadron
state so the combination appearing in the right hand side of (\ref{eq:mups})
can be determined using the known $\Upsilon$ mass and the value of 
$E_{sim}^{\Upsilon}$ which is calculated to be $0.2624(7)$. Then 
\begin{equation}
M_{\Omega_b}=E_{sim}^{\Omega_b}+{\frac{1}{2}}(M_{\Upsilon}-E_{sim}^{\Upsilon})
\label{eq:momegb}
\end{equation}
and 
\begin{equation}
M_{\Omega_{bb}}=E_{sim}^{\Omega_{bb}}+M_{\Upsilon}-E_{sim}^{\Upsilon}.
\label{eq:momegbb}
\end{equation}
The values obtained in this way are given in table \ref{tab:2}. Note that the 
calculated value for the mass of the ground state of $\Omega_b$ is compatible 
with the experimental value \cite{pdg} of $6048(3) MeV$.

In table \ref{tab:2} the results from several recent quark model calculations
are also shown. The quark models have the feature that the energy of the lowest 
lying radial excitation of $\Omega_{bb}$ is significantly smaller than that of 
$\Omega_b$. This is due to different modes being excited as illustrated very
nicely in ref. \cite{Roberts:2007ni} for example. The lattice results also
indicate a reduction of excitation energy in $\Omega_{bb}$ compared to 
$\Omega_b$ but not to the same level as the quark models. A similar effect is
present in a lattice calculation done for charm baryons \cite{Bali:2015lka} 
where, for example, radial excitation energies of $\Omega_c$ and $\Omega_{cc}$
were found to be $626(48)(53) MeV$ and $434(32)(33) MeV$ respectively.
For comparison, the quark model of ref. \cite{Roberts:2007ni} gives
$434 MeV$ for $\Omega_c$ and $365$ for $\Omega_{cc}$.

In this work, and in other lattice QCD studies where single-bottom baryon 
\cite{Woloshyn:2016cgd} and 
heavy-light charm baryon \cite{Bali:2015lka,Padmanath:2014bxa,Padmanath:2015jea}
radial excitation energies were calculated, it is a 
consistent finding that lattice QCD yields excitation energies larger than
quark models. In \cite{Woloshyn:2016cgd} it is also pointed out that 
lattice QCD values for radial excitation energies of singly-heavy baryons
are generally larger or comparable to the excitation energies of heavy-light
mesons while quark model baryonic excitations are found to be smaller than
those of mesons. It should also be noted that lattice QCD and quark models
give reasonably compatible results for excitation energies of 
heavy-light mesons \cite{Woloshyn:2016cgd}. 

It is natural to ask if this work and other the lattice calculations done
to date \cite{Woloshyn:2016cgd,Bali:2015lka,Padmanath:2014bxa,Padmanath:2015jea}
indicate an irreconcilable difference with quark models for heavy-light
baryon spectroscopy. It is probably premature to conclude this as there are
still systematic effects that have not been studied systematically. Lattice
calculations have been done in a variety of lattice setups for up and down 
quark masses, including the present work where there are no valance u,d
quarks and the sea u,d quark masses are near physical. Since the qualitative
comparison of lattice simulations with quark models is the same for all
setups light quark mass extrapolation may not be a primary issue. Perhaps 
more significantly no continuum extrapolation was carried out
in the lattice calculations mentioned above. It is tempting to point to this 
systematic as the source of all discrepancies. However, ground state mass 
values tend to be well described by these simulations so finite lattice 
spacing is not an obvious big issue.

The spatial lattice size used in the simulations discussed here ranges from
$1.8$ to $2.9 fm$. Only in \cite{Bali:2015lka} was the lattice volume
effect considered. For excited states, it was difficult to see the finite 
volume effect due to large statistical fluctuations. What we propose here 
is that differences seen between present lattice simulations and quark models
for radial excitation energies may at the same time be pointing to the 
resolution of the issue.
The lattice QCD discrepancy with quark models is seen to be larger for 
singly heavy baryons than for doubly heavy baryons. Since it is  
expected that singly heavy baryons are more strongly affected by finite 
volume effects this may be indicating the need for baryon simulations in 
larger spatial volumes.

\section{Summary}
\label{sec:3}

Masses for spin ${\nicefrac{1}{2}}$ $\Omega_{b}$ and $\Omega_{bb}$ baryons
were calculated using lattice QCD. As well, the energies of lowest lying 
radial excitations were computed. At present, there is no experimental 
information about radial excitations in heavy-light baryons so comparison
is made to quark models. Quark models have the interesting feature that
the radial excitation energies in doubly-heavy baryons are significantly
smaller than for singly-heavy baryons. This is particularly evident for
bottom baryons. This effect is seen in the present calculation but not
to the extent predicted by quark models.

Summarizing the comparison of this work and a number of other lattice 
QCD calculations 
\cite{Woloshyn:2016cgd,Bali:2015lka,Padmanath:2014bxa,Padmanath:2015jea}
for heavy-light baryons with quark model results one notes that excitation
energies are consistently larger in the lattice calculations. This is
particularly evident for charm baryons \cite{Bali:2015lka,Padmanath:2014bxa}.
However, it is probably premature to conclude that lattice QCD and quark models
predict very different heavy-light baryon spectroscopy since systematic effects
in the lattice simulations still have not been explored fully. The largest 
discrepancy between lattice simulations and quark models occurs for 
single charm baryons (See ref. \cite{Woloshyn:2016cgd}) and decreases for
double charm and bottom baryons. It is suggested that this pattern maybe due
to finite lattice volume effects.

We hope that this work will provide motivation for further studies of 
heavy-light baryons to fill in 
the gaps in lattice QCD simulations and in experimental information.

We thank R.~Lewis and D.~Mohler for helpful comments and the PACS-CS
Collaboration for making their dynamical gauge field configurations
available. TRIUMF receives federal funding via a contribution agreement 
with the National Research Council of Canada.


\begin{thebibliography}{99}

\bibitem{Woloshyn:2016cgd}
R.~M.~Woloshyn and M.~Wurtz,
arXiv:1601.01925[hep-ph].

\bibitem{Roberts:2007ni}
W.~Roberts and M.~Pervin,
Int. J. Mod. Phys. A {\bf 23}, 2817 (2008).

\bibitem{Ebert:2002ig}
D. Ebert, R.~N.~Faustov, V.~O.~Galkin, and A.~P.~Martynenko,
Phys. Rev. D {\bf 66},  014008 (2002).

\bibitem{Yoshida:2015tia}
T.~Yoshida, E.~Hiyama, A.~Hosaka, M.~Oka, and K.~Sadato,
Phys. Rev. D {\bf92}, 114029 (2015).
 
\bibitem{pacscs09}
S.~Aoki, {\emph {et al}.} [PACS-CS Collaboration],
Phys. Rev. D {\bf 79}, 034503 (2009).


\bibitem{Lang:2014yfa}
D.~Mohler, C.~B.~Lang, L.~Leskovec, S.~Prelovsek, and R.~M.~Woloshyn,
Phys. Rev. Lett. {\bf 111}, 222001 (2013).


\bibitem{Lepage:1992}
G.~P.~Lepage, L.~Magnea, C.~Nakhleh, U.~Magnea, and K.~Hornbostel,
Phys. Rev. D {\bf 46}, 4052 (1992).


\bibitem{Lewis:2009}
R.~Lewis and R.~M.~Woloshyn
Phys. Rev. D {\bf 79}, 014502 (2009).

\bibitem{Lewis:2011}
R.~Lewis and R.~M.~Woloshyn
Phys. Rev. D {\bf 84}, 094501 (2011).

\bibitem{pdg}
K.~A.~Olive {\emph {et al}}. (Particle Data Group), 
Chin. Phys. C {\bf 38}, 090001 (2014).

\bibitem{Belle:2012}
R.~Mizuk {\emph {et al}.} (Belle Colaboration),
Phys. Rev. Lett. {\bf 109}, 232002 (2012).
	
\bibitem{Gray:2005}
A. Gray, I. Allison, C.~T.~H.~Davies, E.~Dalgic, G.~P. Lepage, J. Shigemitsu,
and M. Wingate,
Phys. Rev. D {\bf 72}, 094507 (2005).

\bibitem{Davies:2010}
C.~T.~H.~Davies, E.~Follana, I.~D.~Kendall, G.~P. Lepage, and C.~McNeile,
Phys. Rev. D {\bf 81}, 034506 (2010).


\bibitem{Luscher:1990}
M.~L\"uscher and U. Wolff
Nucl. Phys. {\bf B339}, 222 (1990).

\bibitem{Michael:1985}
C.~Michael,
Nucl. Phys. {\bf B259}, 58 (1990).

\bibitem{Blossier:2009}
B.~Blossier, M.~Della~Morte, G.~von~Hippel, T.~Mendes, and R.~Sommer,
J. High Energy Phys. 04, 094 (2009).

\bibitem{Lepage:2001ym}
G.~P.~Lepage, B.~Clark, C.~T.~H.~Davies, K.~Hornbostel, P.~B.~Mackenzie,
C.~Morningstar, and H.~Trottier,
Nucl. Phys. Proc. Suppl. {\bf 106}, 12 (2002).

\bibitem{Ebert:2011kk}
D.~Ebert, R.~N.~Faustov, and V.~O.~Galkin, 
Phys. Rev. D {\bf 84}, 014025 (2011).

\bibitem{Garcilazo:2007eh}
H.~Garcilazo, T.~Vijande, and A.~Valcarce,
J. Phys. G {\bf 34}, 961 (2007).


\bibitem{Bali:2015lka}
P.~P\'erez-Rubio, S.~Collins, and G.~S.~Bali,
Phys. Rev. D {\bf 92}, 034504 (2015).

\bibitem{Padmanath:2014bxa}
M.~Padmanath, R.~G.~Edwards, N.~Mathur, and M~J.~Peardon,
PoS LATTICE2014 084 (2015).
	
\bibitem{Padmanath:2015jea}
M. Padmanath, R.~G.~Edwards, N.~Mathur, and M.~Peardon,
Phys. Rev. D {\bf 91}, 094502 (2015).

\end{thebibliography}
\end{document}